# The Universal Dynamic Complexity as Extended Dynamic Fractality: Causally Complete Understanding of Living Systems Emergence and Operation*

A.P. Kirilyuk

Institute of Metal Physics, Kiev-142, Ukraine 03142
e-mail: kiril@metfiz.freenet.kiev.ua

**Summary.** The unreduced description of any real system with interaction reveals natural emergence of multiple, incompatible system versions, or realisations, formed by dynamic entanglement of the interacting system components and forced to permanently replace each other in a causally random order, which permits one to define and calculate the *a priori* probabilities of their emergence, their internal structure, and the unreduced system complexity. The dynamically redundant and internally entangled realisations form the hierarchical, dynamically probabilistic and self-developing structure of any real system called extended dynamical fractal. Contrary to the conventional fractality, it gives the exact representation of any kind of structure/behaviour, including the specific living system properties, and therefore provides, starting from a high enough level of its complexity, the causally complete understanding of the phenomenon of life and qualitatively new applications in biology and medicine.

## 1 Introduction

The observed forms of many real, and especially living, objects cannot be consistently described in terms of the classical geometry elements, such as lines, spheres or polyhedrons, but should instead be explained as 'infinitely' fine structured, irregular and 'arborescent' (multi-scale) constructions. This fundamental conclusion qualitatively changing the millennia-old ideas of the mathematical presentation of reality underlies the decades-old concept of fractality (see e. g. [1-5]) and marks the end of apparently prosperous development of the canonical science.

However, as it often happens at the beginning of a revolutionary change, the appearing new knowledge inherits, in its starting version, some basic limitations of the previous level. The original notion of fractal, having already become 'canonical' itself and found versatile applications, deals rather with formal, mathematical imitation of the *external* form of *certain* real structures, trying to reproduce their apparently 'self-similar' repetition within a limited range of scales and leaving aside the real mechanism of their autonomous, dynamic emergence, as well as the origin of the whole, unified diversity of forms, including the apparently 'non-fractal' ones. Indeed, it is clear that abstract algorithms used in computer simulations of growing fractals are quite different from any real dynamics by their 'external', 'man-made' origin involving huge simplification of the unreduced interactions governing the natural structure emergence.

As a result, the canonical, 'mathematical' fractals, despite many visible correlations with their natural counterparts, show fundamental and 'fatal' deviations from the essential properties of the latter, such as the totally autonomous and dynamically continuous (though quite uneven) emergence from qualitatively different interacting components, the

---

'true', dynamic randomness (unpredictability) in almost any detail of a natural structure, and global, irregular *intermittency* of 'apparently fractal' and 'apparently non-fractal', 'classical' forms emerging within dynamically *unified* development of the underlying interaction processes (like different structures/organs within a living organism, civilisation, or cosmic space). The latter property of the natural structures means that in reality they do *not* possess scale invariance, apart from an approximate self-similarity existing within certain, fixed group of neighbouring scales and unpredictably passing into quite different type of structure that does not show any self-similarity within the respective group of scales. Besides many evident examples of particular structures, like those cited above, this definitely refers to the ultimately big, 'global' type of meta-structure like a living organism (considered now at *all* levels of its dynamics), the planet ecosystem, or the whole universe that should most probably emerge, at least within the majority of the scales involved, by a dynamically continuous development of the naturally incorporated interaction processes giving thus the single, unified dynamical fractal of the 'global' system that includes the whole diversity of its structures.

The non-dynamical, 'kinematic' or 'geometric', nature of the canonical fractals determines not only their fundamental difference from the real prototypes, but also the actual absence of the common basis between the 'fractal' presentation of reality and other views on the same reality obtained within various, also basically limited and separated among them, concepts of the canonical 'science of complexity' (such as 'dynamical chaos', 'self-organisation', 'adaptability'), despite many occasional links between them (cf. [6]). A related basic deficiency of the usual fractality, and 'science of complexity' in general, is that its approach to higher-level, living systems is basically detached from explanation of the world dynamics at other, lower levels that determine, for example, the operation of microscopic physical and chemical constituents of a living organism (some of them are even not considered as complex systems and fractals within the canonical 'science of complexity'). Since in addition the scholar science presentation of that, relatively 'simple' kind of systems is characterised itself by the fundamental incompleteness and persisting difficulties with causal understanding of reality ('quantum mysteries' etc.), it becomes clear that the causally complete, fundamentally consistent description of the resulting, much more complex biological systems, imposed by the modern stage of development, is possible only within essential, qualitative extension of the canonical science paradigm.

It is especially important to have the unreduced, dynamically based understanding of fractality when it is related to living organisms, since they are just characterised by the explicit, even externally 'strong' manifestations of the full-scale autonomous creativity ('spontaneous' emergence of structures, permanent autonomous motion at all levels, and 'desire' of change, 'élan vital'), dynamical randomness, *unified* diversity (wholeness), extremely large adaptability, 'individuality', etc. Indeed, it is clear that any canonical, mathematically or numerically simulated fractal that does not possess all those essential properties in their unreduced version cannot reproduce the major system property of 'being alive', after which any its partial, external similarity to an artificially fixed form of the real organism can only be useful for lower-level, 'mechanical' and empirically based applications, but not for the intrinsically complete understanding and monitoring of life becoming absolutely indispensable at the modern stage of quantitatively powerful technology development. In practice, the canonical fractality operates with *statistically averaged*, external (geometro-kinematic) characterisation of the observed structures and their numerical simulations (like 'fractal dimensions' etc.) [1-5], whereas it is a variously confirmed fact that all most specific features of living forms, and the essence of the phenomenon of life as such, are determined just by non-statistical, individually specified, 'inimitable' structures and types of behaviour.

In this paper we present the causally complete extension of the notion of fractality totally adequate with respect to real (including living) systems it describes and obtained within the new, universal concept of dynamic complexity based on the phenomenon of *dynamic multivaluedness (redundance)* of any interaction process [7]. Using these results we show (section 2) that such *extended dynamical fractals* reproduce exactly the mentioned essential properties of real (living) systems and therefore provide the correct and universal representation of any local or 'global' structure, including its apparently 'non-fractal' parts (in terms of canonical fractality). This means that the unreduced dynamical fractal is none other than the exact structure of the hierarchy of complexity representing the *whole* world dynamics and taken at any its particular location or group of levels, where living systems (in the ordinary sense) start from certain, high enough level of that unified, fractal hierarchy of dynamic complexity of the universe [7]. This inbred universality and unreduced realism of the extended fractality open unrestricted, qualitatively new possibilities for various practical problem solution in biology and medicine outlined in section 3.

## 2 Universal dynamic fractality within the dynamic redundance paradigm and the properties of life

We start with a rather evident, and actually well substantiated, argument that any real system can exist and possess its properties only due to the interaction between its elementary constituents (coming eventually from lower levels of the same, unified hierarchy of world complexity [7]). It is equally obvious that all real interactions are 'unreduced' in the sense that within a system 'everything interacts with everything'. However, since such unreduced interaction does not allow for a closed, 'exact' solution, the canonical science, including all approaches of the conventional 'science of complexity', invariably resorts to essential truncation of self-developing interaction links within a version of 'perturbation theory' transforming the problem into a 'separable', or 'integrable' one (possessing an exact solution), but simultaneously killing just those non-trivial properties of the natural interaction that determine the emergence and development of the real system. It is especially evident that the unreduced, self-developing and dynamically unpredictable fractality cannot be obtained as a 'closed' solution of a 'separable' problem.

Therefore we try to analyse the arbitrary interaction process within a real system without any simplification, which leads, as we shall see, to the discovery of its qualitatively new properties, such as dynamic randomness, fractality and intrinsic creativity. We can only briefly illustrate here the rigorous analysis of the universal science of complexity [7] that starts with a quite general mathematical representation of a system with interaction by its 'existence equation' simply fixing the fact of interaction. Consider for simplicity and without loss of generality interaction between two real entities, Q and Ξ (they can be subdivided into any number of independent constituents), characterised by the degrees of freedom $q$ and $\xi$. In course of interaction the component degrees of freedom inseparably entangle among them and produce a compound system state described by the *state-function*, $\Psi(q,\xi)$, which satisfies the following existence equation simply describing the conditions of the problem:

$$\left[h_Q(q) + h_\Xi(\xi) + V(q,\xi)\right]\Psi(q,\xi) = E\Psi(q,\xi) , \qquad (1)$$

where the 'generalised Hamiltonians' $h_Q(q)$ and $h_\Xi(\xi)$ describe the 'free' state of interacting entities 'before' the beginning of interaction (they can be represented by any actually measured function, being a form of 'dynamic complexity' defined below, but can always be interpreted as Hamiltonians in the universal formalism [7]), $V(q,\xi)$ is the generalised 'interaction potential' bringing inseparability of entanglement to the system, and $E$ is the 'eigenvalue', corresponding to the chosen 'Hamiltonians' and expressing the measured property value for the whole compound system.

Instead of trying to artificially ('approximately') separate the irreducibly entangled interacting entities by developing a version of 'perturbation theory' (conventional approach), we can trace the real interaction development by recurrently expressing one of the entities (say $q$) through the other one ($\xi$), using eq. (1), and thus reformulate the existence equation in terms of only one entity ($\xi$), while the other entity (system components) will actually be present in the total solution reflecting the recurrent development of interaction links (component entanglement). We thus obtain the *effective existence equation* first used for physical problems solution [8,9] and generalising the well-known 'optical', or effective, potential method:

$$\left[h_\Xi(\xi) + V_{\text{eff}}(\xi;\eta)\right]\psi_0(\xi) = \eta\psi_0(\xi) , \tag{2}$$

where the operator of the *effective (interaction) potential (EP)*, $V_{\text{eff}}(\xi;\eta)$, is given by

$$V_{\text{eff}}(\xi;\eta) = V_{00}(\xi) + \hat{V}(\xi;\eta), \quad \hat{V}(\xi;\eta)\psi_0(\xi) = \int_{\Omega_\xi}d\xi' V(\xi,\xi';\eta)\psi_0(\xi') , \tag{3a}$$

$$V(\xi,\xi';\eta) \equiv \sum_{n,i} \frac{V_{0n}(\xi)\psi_{ni}^0(\xi)V_{n0}(\xi')\psi_{ni}^{0*}(\xi')}{\eta - \eta_{ni}^0 - \varepsilon_{n0}} , \quad \varepsilon_{n0} \equiv \varepsilon_n - \varepsilon_0 , \tag{3b}$$

$$V_{nn'}(\xi) \equiv \int_{\Omega_q}dq\,\phi_n^*(q)V(q,\xi)\phi_{n'}(q) , \tag{3c}$$

$\varepsilon_n$ and $\phi_n(q)$ are the eigenvalues and eigenfunctions for the non-interacting entity/object Q describing its elementary dynamical 'modes' or (physical) 'points', $\eta \equiv E - \varepsilon_0$ is the eigenvalue of the unreduced problem to be found, $\{\psi_{ni}^0(\xi)\}$ and $\{\eta_{ni}^0\}$ are the complete sets of eigenfunctions and eigenvalues for an 'auxiliary' system of equations,

$$\left[h_\Xi(\xi) + V_{nn}(\xi)\right]\psi_n(\xi) + \sum_{n'\neq n}V_{nn'}(\xi)\psi_{n'}(\xi) = \eta_n\psi_n(\xi) , \quad \eta_n \equiv E - \varepsilon_n , \tag{4}$$

that just describes the 'outlet' to further development of inseparable (nonintegrable) interaction process, and summations are performed over $n,n' \neq 0$.

The total solution of eq. (1) is expressed as

$$\Psi(q,\xi) = \sum_i c_i \left[\phi_0(q) + \sum_n \phi_n(q)\hat{g}_{ni}(\xi)\right]\psi_{0i}(\xi) , \tag{5a}$$

$$\hat{g}_{ni}(\xi)\psi_{0i}(\xi) = \int_{\Omega_\xi}d\xi' g_{ni}(\xi,\xi')\psi_{0i}(\xi') , \quad g_{ni}(\xi,\xi') = \sum_{i'} \frac{\psi_{ni'}^0(\xi)V_{n0}(\xi')\psi_{ni'}^{0*}(\xi')}{\eta_i - \eta_{ni'}^0 - \varepsilon_{n0}} , \tag{5b}$$

where the eigenfunctions, $\{\psi_{0i}(\xi)\}$, and eigenvalues, $\{\eta_i\}$, are found from the effective existence equation, eqs. (2)-(3), and coefficients $c_i$ should be determined from solution matching on the boundary where $V(q,\xi) = 0$. The measured generalised system density, $\rho(q,\xi)$, is given either by the state-function itself, $\rho(q,\xi) = \Psi(q,\xi)$, or by its simple function like squared modulus, $\rho(q,\xi) = |\Psi(q,\xi)|^2$.

The 'effective' formulation of a problem, eqs. (2)-(5), is formally equivalent to its initial expression, eq. (1), but now it permits one to explicitly reveal the universal mechanism of the *essential nonlinearity* development within *any* unreduced interaction process (which may seem externally linear, in the initial formulation of eq. (1)). Namely, the essential nonlinearity enters through the EP dependence, eqs. (3), on the eigenvalues to be found, $\eta$, which does not reflect any *a priori* presumed, artificially inserted and arbitrary defined 'nonlinearity' of the potential configuration or its relation to the state-function, but emerges *dynamically*, in course of the natural interaction development always involving, in its unreduced version, the autonomous interaction loop formation ('everything interacts with everything' and through everything with itself) just adequately described by the recurrent EP formalism.

The essential, dynamic nonlinearity is eliminated in the perturbative, 'separating' approach of the canonical science that tries to 'simplify' the exact expressions of eqs. (3)-(5) by totally reducing the EP dependence on the unknown eigenvalues and eigenfunctions and obtains indeed an 'exact', closed solution, but only in the form of small, perturbative additions to the zeroth order problem (obtained, for example, if $V_{\text{eff}}(\xi;\eta) = V_{00}(\xi)$ in eq. (3a)). After that the conventional science often inserts an artificial, but also integrable 'nonlinearity' which in reality is equivalent, in terms of the real, dynamical nonlinearity, to another *essentially linear* problem that corresponds to 'renormalisation' of the 'free', separated system state expressed now through other, as if 'nonlinear', but in reality always 'exact' (integrable), eigen-modes.

If now we preserve the essential nonlinearity of the unreduced EP formalism, eqs. (2)-(5), then it appears that the solution to a problem can still be found, but it possesses qualitatively new properties, providing the described system behaviour with the intrinsic, dynamical randomness and self-developing, unreduced fractality. Indeed, it is not difficult to see [7-9] that the nonlinear appearance of the eigenvalue to be found, $\eta$, in the EP expression, eqs. (3), leads to excessive number of eigen-solutions (with respect to their ordinary 'complete' set) found from the characteristic equation, since the additional dependence on $\eta$ increases its highest power in that equation. If $N_\xi$ is the number of terms in the sum over $i$ in eq. (3b), actually equal to the number of eigen-modes (or 'points') of the object/entity $\Xi$ participating in the interaction, then one obtains the $N_\xi$-fold redundance of the eigen-solutions of the effective existence equation, eq. (2), where each of the redundant solutions is as complete as any ordinary solution and exhaustively describes a state of the system. Since all these solutions, representing different 'versions' of the system and therefore called system *realisations*, are absolutely equal in their 'right' to appear in observations, the system is forced to permanently 'jump' from one realisation to another, each time passing by a special 'intermediate' state/realisation and 'choosing' each next 'regular' realisation from their full set in a *causally random* fashion. This means that every smallest, even infinitesimal (or 'fractal') fluctuation of the system in the intermediate state towards a particular realisation is dynamically self-amplified and the system undergoes a catastrophic 'collapse', or 'reduction', to this realisation, after which it returns to the intermediate realisation and the cycle repeats with a new, causally random realisation choice. This *unceasingly* repeated *dynamical reduction* of the system to a randomly chosen realisation and the reverse *dynamic extension* to the common (delocalised) intermediate realisation are consistently described by the above equations of the unreduced EP formalism [7]. Thus, it can be seen from expression of eq. (5b) for the state-function that the latter is concentrated (dynamically squeezed) just around that configuration/point, where the corresponding EP well is dynamically formed, according to eq. (3b), for that particular realisation.

We call the revealed multitude of incompatible (redundant) realisations of an arbitrary system with interaction *dynamic multivaluedness (redundance)* phenomenon and show that it is the key feature of the *universal dynamic complexity* of any real system (interaction process), where dynamic complexity as such can be consistently defined as any growing function of the total number of system realisations, or the related rate of their change, equal to zero for the unrealistic limiting case of only one system realisation [7,9]. It is that, artificial projection of the dynamically multivalued reality onto single, 'averaged' solution-realisation that is exclusively studied in the canonical science, including the conventional 'science of complexity', which provides the dynamically single-valued, effectively one-dimensional, and therefore regular (fixed), imitation of the 'living', dynamically probabilistic realisation change process.

Being rigorously derived from the unreduced analysis of a generic interaction process, the dynamic redundancy phenomenon has also a transparent physical interpretation: the unreduced interaction between two entities with $N$ elements each gives $N^2$ versions of their dynamical 'contacts' ('everything interacts with everything'), while the number of 'places' in reality for the interaction products is the same as that for the free interaction participants, and equals $N$, which gives $N$-fold redundancy of interaction products. The resulting causal randomness of realisation change process can be mathematically expressed in the form of extended, now really complete *general solution* of a problem obtained as the *causally probabilistic* sum of the measured system densities, $\rho_r(q,\xi)$, for all realisations:

$$\rho(q,\xi) = \sum_{r=1}^{N_\Re} {}^\oplus \rho_r(q,\xi) , \qquad (6)$$

where the sum is performed over the total number $N_\Re$ ($\leq N_\xi$) of the actually observed (discerned) system realisations, the generalised system density for the $r$-th realisation $\rho_r(q,\xi)$ is obtained from the system state-function expression, eqs. (5), with the corresponding eigen-solution set for this realisation, $\{\eta_i^r, \psi_{0i}^r(\xi)\}$, found from eqs. (2)-(3) and substituted for their general notation $\{\eta_i, \psi_{0i}(\xi)\}$, and the sign $\oplus$ serves to designate the special, dynamically probabilistic meaning of the sum. This implies, in particular, that the probability, $\alpha_r$, for the $r$-th realisation emergence can now be determined *a priori* (theoretically), by the unreduced analysis of system dynamics that explicitly *derives* the *complete* set of incompatible *events/realisations*. Since all the emerging solution-realisations have absolutely equal 'status' of really existing system states *forced to emerge* by the driving interaction, it is evident that $\alpha_r = 1/N_\xi$ for any individual, 'elementary' realisation and $\alpha_r = N_r/N_\Re$ for a 'composite' realisation containing within it $N_r$ experimentally unresolved elementary realisations. Correspondingly, one can discern two characteristic, opposite cases of dynamically multivalued (complex) behaviour determining the whole diversity of observed patterns between them [7]: the regime of *uniform chaos* is characterised by sufficiently different realisations with approximately equal probabilities (very irregular, distributed kind of behaviour), whereas the regime of (multivalued) *self-organisation*, or *self-organised criticality (SOC)*, emerges when one has a number of groups of sufficiently similar realisations (within each group) and quite inhomogeneous distribution of realisation probabilities (externally distinct, though *always* intrinsically probabilistic, structures). In any case, the expectation value of generalised system density, $\rho_{\rm ex}(q,\xi)$, measured as average over a large number of equivalent events is obtained as

$$\rho_{\rm ex}(q,\xi) = \sum_{r=1}^{N_\Re} \alpha_r \rho_r(q,\xi) , \qquad (7)$$

with the *dynamically*, *a priori* determined, probabilities $\{\alpha_r\}$ (they are actually related to the coefficients $c_i$ in the state-function expression, eq. (5a), found from dynamic solution 'matching' in the intermediate realisation phase between each two successive regular realisations [7,9]). Note, however, that in the general case described by eq. (6) one can determine from the first principles unpredictable system variations and their probabilities also for rare, *individual* events of realisation emergence.

Another key feature of unreduced interaction complexity inseparably related to causal randomness of realisation change is the autonomous, *dynamic entanglement* of the interacting entities within each realisation. In the above physical picture of the dynamic redundance phenomenon we have *N* redundant versions (realisations) of physically real entanglement, or 'engagement', between the elements (modes, points) of interacting entities realising their particular dynamical 'mixing' within the emerging compound entity of the thus formed new 'level of complexity' of the world. This dynamic entanglement specifying, together with the probabilistic realisation change, the real sense of the 'interaction process' is described by the products of functions of *q* and *ξ* in the state-function (density) expression, eq. (5a), and is reflected also in EP formalism expressions containing matrix elements $V_{nn'}(\xi)$, eq. (3c). Dynamic entanglement is an aspect of the intrinsic dynamic instability of any real interaction described by its essential nonlinearity and leading to unceasing series of catastrophic system collapses (reductions) to its consecutively emerging realisations. During each collapse event the interacting entities increasingly interpenetrate in an avalanche-type, self-amplifying 'fall' until a transient 'equilibrium' of the completely formed realisation is attained, and then the opposite, equally self-accelerating *dynamic disentanglement* phase begins returning the system to the disentangled, 'quasi-free' state of the intermediate realisation immediately followed by another dynamic entanglement-collapse towards configuration of the next, randomly chosen realisation, and so on.

It is at this point that the *extended dynamic fractality* naturally emerges as unavoidable continuation of the dynamic entanglement at progressively finer scales due to the 'inseparable' character of the interaction process (now also in a well-specified physical sense) reflected in the essential use of the unknown eigen-solutions, $\{\eta_{ni}^0, \psi_{ni}^0(\xi)\}$, of the auxiliary system of equations, eqs. (4), in the main state-function and EP expressions of the 'effective' formalism. Since the essential nonperturbative manifestations of unreduced interaction, the dynamic redundance and entanglement, are *already* revealed and *exactly* taken into account by the particular structure of the EP formalism, we can use a suitable approximation (e. g. of a 'mean-field' type) to find a closed expression for these eigen-solutions from eqs. (4) [7-9], as far as manifestations of this particular level of complex dynamics are involved. However, there is always a link to finer levels of the same system dynamics, and these can be obtained by using the same EP analysis for unreduced solution of the auxiliary system, eqs. (4). We thus obtain the same dynamic redundance (causal randomness) and entanglement phenomena, but now on that new, finer scale which, in its turn, is dynamically related to still finer levels of dynamical splitting and entanglement (we do not explicitly cite here the corresponding equations obtained similarly to those of the above formalism [7]). The dynamically fractal, arborescent internal structure of the unreduced interaction process emerges thus as a (theoretically) infinite, internally continuous, but quite uneven (dynamically discrete, or 'quantized'), sequence of ever finer levels (scales) of probabilistically changing realisations, each of them containing the unceasing processes of dynamic entanglement/disentanglement of interacting entities at its own level and all finer levels. This quite involved and 'living', everywhere *pulsating* hierarchy of real interaction dynamics is quite different from its over-simplified, single-valued (averaged), rigidly fixed projection onto one-dimensional presentation of the conventional perturbative analysis and expresses the fact that the resulting extended dynamical fractal can be described as really complete *general solution* to an *arbitrary* problem (of any system dynamics).

The qualitatively extended character of this causally complete structure of real dynamical fractal, consistently derived from the unreduced interaction analysis, with respect to the canonical fractality is obvious and reduced to the following main features: unceasing, causally probabilistic *motion* of emerging 'branches' (representing, in general, denser groups of similar elementary realisations); their well specified physical 'texture' (material 'quality') obtained from the (multilevel) dynamic entanglement of interacting

4entities; and the dynamically determined, permanent self-development of the unreduced fractality. It is this latter property and its direct relation to the preceding ones (dynamically redundant entanglement) that determine the intrinsically 'living' character of the extended dynamical fractal and are especially important for its biomedical applications (section 3). In particular, since dynamical splitting into growing branches/realisations is driven exclusively by the probabilistically emerging interaction configurations, the detailed appearing structure takes the 'final' form that automatically fits in with the 'environment' and within itself: this is the extended, *dynamic adaptability* of real complex systems, including living organisms (and their evolution), only mechanistically, inconsistently simulated within the artificially imposed schemes of the conventional, single-valued science and 'complexity' concepts. The sequence of levels of real fractal branching is subjected to the same dynamic unpredictability of details as splitting into incompatible realisations (and their groups) at each level of fractality, which can give the (always partially broken) regularity of the canonical fractal 'scale symmetry' only within a limited, occasionally occurring group of scales (they correspond to a denser group of similar system realisations within the 'self-organisation' regime of multivalued dynamics, defined above). This means also that the unreduced fractal naturally, *dynamically* produces *all* types of structures/objects and behaviours, including the apparently 'non-fractal', 'solid' ones. As for the internal structure of fractal branches described by the probabilistic dynamic entanglement of interacting entities in the unreduced science of complexity, it is clear that conventional simulations usually do not even try to analyse it because of their basically external, non-dynamical approach.

    The self-maintained, locally unrestricted *growth* of the extended dynamical fractal as a result of development of any system with interaction can be described by the universal law of *conservation, or symmetry, of complexity* which is realised through the unceasing dynamic *transformation* between two basic forms of dynamic complexity, the *dynamic information* and (generalised) *dynamic entropy* [7]. Both dynamic information and entropy are expressed through the same, suitable measure of complexity (always eventually related to system realisations number or rate of their change, see above), but (physically real) information represents the stock of system complexity in a 'hidden'/latent, or 'folded', form at the beginning of interaction process (it is a generalised 'potential energy' of interaction), while the dynamic entropy is the explicit, 'unfolded' form of the same complexity characterising the final result of interaction development (at a current level), such as the generalised 'kinetic/thermal energy' of the appeared object or its 'structural complexity'. Therefore the system dynamic information, $I$, always decreases in course of system interaction development (dynamical fractal emergence), while the dynamic entropy, $S$, always increases, but since the increase of $S$ is simply the result of the transformation of $I$, their sum, the *total system complexity C*, remains unchanged: $C = I + S =$ const, $\Delta S = - \Delta I$. In this way, the universal conservation/symmetry of complexity provides also the universal system evolution law that can be specified in the form of the universal Hamilton-Lagrange and Schrödinger equations causally related within the single unified formalism and taking the form of any known (correct) equation or underlying conservation law (or any other basic 'principle' empirically postulated in the canonical science), in the corresponding particular situation [7].

    These rigorously derived results of the universal science of complexity permit us to avoid the unpleasant 'bias' and inconsistency of the canonical 'law of entropy growth' by presenting the generalised dynamical entropy growth as an inevitable consequence of conservation or *symmetry* of complexity (which, in addition, is consistently substantiated). This is related also to the fact that the dynamical entropy growth is associated not only with some visible degradation ('growth of disorder'), as it is implied by its canonical concept, but accompanies *every* process, including the emergence of apparently 'ordered' structures (e. g. living organisms) inconsistently interpreted by the canonical science as (actually arbitrary large) 'fluctuations' on the background of 'general degradation'. This difference can be understood in terms of the dynamic redundance

phenomenon: any *real* structure, however 'ordered' or 'static' it may seem, is always maintained due to the *dynamically random* (causally probabilistic) order of *different* realisation *unceasingly* changing within it, even though those realisations can be quite similar (but never identical) to each other. It is clear why this causal randomness, and the related internal 'disorder', cannot be obtained within any dynamically single-valued, conventional form of knowledge.

The universal symmetry of complexity is quite different from any 'exact', 'superposition' symmetry, postulated in the canonical science, by its dynamical, 'emergent' origin and character. The latter leads, on one hand, to the above universal evolution law, including its causally understood 'sense' (irreversible direction) of growing complexity-entropy (decreasing complexity-information). On the other hand, for any given, already formed structure the symmetry of complexity means the dynamic symmetry between all system realisations (that never stop to replace each other) which necessarily show some, generally quite large and always partially irregular, variations in their configuration. Therefore the universal symmetry of complexity is *always* naturally 'broken', in full accord with observations and in contrast to the invariably 'regular' symmetries of the dynamically single-valued science obliged to introduce (postulate) the notion of 'broken symmetry' artificially, in order to account for the observed 'asymmetric world'. This fundamental difference between complex-dynamical and mechanistic concept of symmetry manifests itself also in the difference between the basically exact 'scale symmetry' (self-similarity) of the canonical fractal and the symmetry of complexity of the extended dynamical fractal that rather *excludes* any exact regularity in the scale transformation result (since it would give unrealistically simple, zero-complexity structure). Only the dynamically multivalued science can state that, and consistently explain why, any living organism, with all its apparent irregularity, is indeed much *more* symmetric (= dynamically complex), than visibly regular (or irregular) 'inanimate', lower-level objects.

The extremely large diversity of extended dynamical fractals unifying and naturally producing both 'fuzzy' and 'solid' type of structures is vitally important for the autonomous, self-sufficient character of complexity development, since it becomes clear that the emerging 'solid' branches of the dynamical fractal (realising the above regime of 'multivalued SOC') play the role of 'interacting entities' for next (higher) level(s) of developing complexity, while the 'infinitely' fine 'foliage' (varying between SOC and 'uniform chaos' regimes) serves as their (tangible) 'interaction' as such determining the next portion of dynamical information ('potential energy') to be transformed into dynamic entropy at that, higher level of complexity unfolding.

## 3 Creative control of life with the help of extended dynamic fractality

It is reasonable to start any fundamental concept application to living systems with the consistent definition, within this concept, of the phenomenon of life as such, including all its distinctive features. We have shown above that the main properties of the unreduced, dynamically multivalued fractality reproduce exactly living organism behaviour. Therefore we can define life as the unreduced dynamic complexity (obtained by the dynamic redundance paradigm), starting from certain, high enough level of its naturally emerging hierarchy. It is also clear now why any attempt to understand the phenomenon of life within the dynamically single-valued paradigm of the canonical science (including the conventional 'science of complexity' and related numerical simulations) cannot be consistent in principle.

In particular, we can see the origin of fundamental deficiency in the definition of life in terms of physics proposed by E. Schrödinger [10] and considering a living organism as a 'deviation' from the conventional law of entropy growth that can appear because the organism is an 'open system' that can actively 'take' (mainly 'eat up') the necessary

'order', or 'negative entropy', from the 'environment'. It is easy to see that such definition cannot be consistent already because, at least for higher organisms, the whole 'negative entropy' consumed from the environment is much lower than the obtained organism complexity (thus, the complex brain dynamics reproduces with high enough accuracy the unreduced complexity of a 'large' environment that considerably exceeds the amount of the consumed nutrition, let alone individual psycho-physiological reactions, emotions, conscious thinking, etc.). The extended, complex-dynamical definition of entropy and related law of conservation-transformation of complexity show that the organism contains the main part of its complexity, in the latent form of dynamical information, already at the beginning of its life and then the information only unfolds itself into explicit form of dynamical entropy, which does not contradict (in contrast to the conventional entropy concept) the high dynamical order of the living organism, since it is but highly inhomogeneous distribution of probabilities of realisations replacing each other in a causally random order. The dynamic information is evidently 'hidden' within the genome, while the matter/energy exchange with the environment, being a necessary component of the organism development, accounts for a smaller part of the resulting complexity-entropy (it is sufficient to compare the amounts of products consumed e. g. by conscious and various non-conscious organisms and their respective complexity levels). The period of life of any individual organism can be described as progressive transformation of its dynamical information (from the genome) into the dynamical complexity-entropy of the adult (and then old) organism, which determines the particular (always finite) duration and 'sense' (or 'character') of any individual life, including the necessary interaction with the environment also driven by the main $I \to S$ transformation process. This complex-dynamical definition of life provides the evident causal interpretation for the 'teleological' ('reasonable') type of behaviour characteristic of living organisms, where the (latent) stock of dynamical information is the causally complete version of the famous 'élan vital' [11] giving rise to artificially mystified speculations about 'life force' within the dynamically single-valued approach.

    We can proceed with applications of the extended concept of fractal complexity in biology and medicine by describing a universal way of practical realisation of the dynamic redundance paradigm for causally complete solution of particular problems based on the fundamental results of section 2. It should generally include the following stages that may use more formalised (mathematical) or more qualitative (but equally well substantiated) form of the universal science of complexity always preserving, however, its irreducible causal content:

(1) find the key, 'driving' *interactions* and interacting *entities* at the current level(s), including both 'natural' objects and 'artificial' means of their modification (in continuous dynamic connection to lower levels of complexity);
(2) perform the *unreduced* interaction analysis that should give the *complete* set of *incompatible*, chaotically *changing* system realisations, including their structure and probabilities (dynamic entanglement and multivaluedness), as well as their fractal hierarchy, where necessary;
(3) identify the realisations with the *emerging* entities and regimes of behaviour;
(4) verify (e. g. by problem variation) the found realisation set for *completeness*;
(5) establish *dynamic* connection to higher levels of system complexity and the *unreduced* complex dynamics of the environment; if necessary, return to (1).

We emphasize that this sequence of stages applies not only to purely theoretical analysis, but also to empirical, semi-empirical, computer investigations and all their combinations.

    Among general biological applications of the extended fractality/complexity we can mention the theory of quasi-autonomous, *dynamically creative evolution* and *causally complete genetics*. The biological (natural) evolution can be understood now as a case of universal complexity unfolding process (multi-level transformation of dynamic information into entropy), which permits one to consistently explain, taking into account

the unreduced complexity properties, such 'difficult' features of evolution as its very uneven and irregular character showing permanent dynamic interplay of order and randomness. Whereas any conventional, single-valued approach, including the classical Darwinism and all its modern modifications, can explain *only selection* among *ready*, though varying, forms, but *not* their natural *emergence*, the unreduced science of complexity provides the universal mechanism of creation of qualitatively new forms (see section 2) and thus explains the direction and sense of evolution as progressively growing (dynamically unfolding) complexity-entropy of the whole biosphere, while particular correspondence between the new species and the 'environment' ('survival of the fittest') need not be very exact or regular and should rather be considered as a natural consequence of the above general 'purpose'. The fractal hierarchy of feedback loops of the unreduced interaction governing new form creation points also to a causal extension of the Lamarckian approach to evolution specified as reverse influence of the level of (now objectively determined) dynamic complexity attained by the organism (including its interaction with the environment) on the dynamic information (structure) of its genome (further theoretic and experimental research is needed for understanding of the details).

Causally complete genetics is obtained as a complex-dynamical (multivalued) extension of the conventional genetics always using a perturbative, single-valued approach and treating genetic information and its transformation into living structures as a one-dimensional, dynamically regular line-program of the ordinary computer ('Turing machine'). It is clear that the genetic machine operation is determined by the natural, 'unreduced' interactions (in the sense specified above) leading to emergence of dynamically multivalued, causally probabilistic forms and having nothing to do with single-valued mechanistic 'copying' and 'sequencing' implied and performed by the canonical genetics. It is important to note that the continuing application of just that huge simplification of the real interaction complexity in intense, technically powerful genetic experimentation can, and inevitably does, lead to practically dangerous, catastrophic consequences of blind manipulations with intrinsically unpredictable system. In order to obtain truly progressive, intrinsically safe and constructive development in this, and various other, fields with direct, profound modification of irreducibly complex systems allowed for by the attained high level of empirical technology, the level of fundamental knowledge should correspond to the unreduced complexity of the real system (as it is presented by the universal general solution of a problem in the form of extended dynamical fractal described above).

*Medical applications* of the unreduced fractal complexity are based on the causally complete understanding of living organism dynamics provided by the extended concept of fractality/complexity and lead to the new stage of *integral medicine* allowing for the creative control/extension of the multivalued organism dynamics guided by the universal criterion of progress of the unreduced science of complexity, i. e. most complete, optimal transformation of complexity-information into complexity-entropy. This approach permits one to escape from the modern impasse of mechanistic empiricism in medicine when a 'complex-dynamical', conscious doctor tries to cure an irreducibly complex, inimitable organism with the help of non-complex, over-simplified, standard means: such single-valued imitation of the highly multivalued dynamics is reduced to superficial suppression of external signs of illness ('treatment of symptoms'), instead of correcting the real, complex-dynamical origin of the disease ('treatment of the disease'). It is also evident that application of the conventional 'science of complexity', falling within the same essentially perturbative paradigm of dynamic single-valuedness ('exact solutions') and thus only *imitating* (or 'modelling') separated *external manifestations* of complexity, cannot really change the situation. In the integral medicine approach the causally complete pattern of the extended fractal of organism dynamics is explicitly known and consciously modified using the basic property of essential nonlinearity/instability of the multivalued dynamics, which provides the universal 'natural' cure unifying, at a superior level, the advantages of 'technological' and 'traditional' tendencies in medicine [7]. Some specific results of this intrinsically creative medicine can take the form of individual multi-dimensional (fractal and changing) 'chart' of organism state as a universal tool for absolute, in principle, personal health control, or unreduced, complex-dynamical (re)construction of living tissues/organisms through the natural growth of the corresponding extended dynamical fractal, or preparation of 'living' medicaments having the unreduced dynamic fractal structure and full complexity of living objects, etc. In each particular application one can use the same, universal approach of the unreduced science of complexity (see items (1)-(5) in this section), which provides the fundamental advantage of the *unified*, naturally developing understanding of various levels and aspects of organism dynamics and allows for natural inclusion into consideration of other (e. g. 'psychological', 'social', or 'ecological') levels of complexity where necessary. In particular, such unified, causally complete understanding permits one to overcome the impasses of the canonical medicine in treatment of 'really difficult' (and often dangerously 'emerging') diseases, such as cancer or AIDS, that simply involve the irreducible complexity levels inaccessible to the conventional empiricism and dynamically single-valued analysis, irrespective of the quantitative power of the instruments used. The qualitatively growing technological power of civilisation necessitates in itself the corresponding up-grade of the conceptual understanding, in order to transform the real danger of modern blind, but critically deep experimentation into unlimited possibilities of the causally complete knowledge.